\documentclass[pre,showpacs]{revtex4}
\pdfoutput=1 
\usepackage{amsmath}
\usepackage{graphicx}
\begin{document}

\title{A generalized voter model on complex networks}

\author{Casey M. Schneider-Mizell, Leonard M. Sander}
\affiliation{Department of Physics, University of Michigan, Ann Arbor, MI 48105}

\date{\today}

\begin{abstract}
We study a generalization of the voter model on complex networks, focusing on the scaling of mean exit time. Previous work has defined the voter model in terms of an initially chosen node and a randomly chosen neighbor, which makes it difficult to disentangle the effects of the stochastic process itself relative to the network structure. We introduce a process with two steps, one that selects a pair of interacting nodes and one that determines the direction of interaction as a function of the degrees of the two nodes and a parameter $\alpha$ which sets the likelihood of the higher degree node giving its state. Traditional voter model behavior can be recovered within the model. We find that on a complete bipartite network, the traditional voter model is the fastest process. On a random network with power law degree distribution, we observe two regimes. For modest values of $\alpha$, exit time is dominated by diffusive drift of the system state, but as the high nodes become more influential, the exit time becomes becomes dominated by frustration effects. For certain selection processes and parameters values, an intermediate regime occurs where exit occurs after exponential mixing.
\end{abstract}


\maketitle

\section{Introduction}

The voter model has been extensively 
studied on lattices~\cite{Liggett05a} and, in recent years, on complex networks  \cite{Suchecki05a,Sood:2005p116,Suchecki:2005p95,Castellano:2005p124, Vazquez:2008p5347} and is closely related to models of language evolution ~\cite{Baxter:2006p482}, ecological dynamics~\cite{PHubbell:2001p959}, opinion dynamics~\cite{Castellano:2007p907}, and epidemic spread~\cite{PastorSatorras:2001p2157}. The voter model defines a dynamical process where nodes are each assigned one of two states, $+1$ or $-1$. Connections are defined on a lattice by nearest neighbors or on an arbitrary network by edges. Each update step consists of selecting a pair of nodes and giving the state of one node to the other. In the most frequently studied version of the model, the first node chosen adopts the state of the second. The most interesting object of study is the mean exit time, i.e. the mean time to achieve complete agreement. For the voter model on complex networks, a node will be chosen second with frequency proportional to its degree, and so its influence is fixed by the selection process. In this paper, we introduce a generalized voter model with a single tunable parameter that allows control of the influence of topology in a manner independent of the selection process. In this generalized model, the probability of a node giving its state to its neighbor is proportional to $k^\alpha$, where $k$ is the node's degree and $\alpha$ is chosen.

Voter model processes based entirely on selection frequency are denoted either \emph{link update} or \emph{node update} \cite{Suchecki:2005p95}. In link update dynamics, every time step an edge is selected uniformly at random. One of the two nodes at the ends of the link is then chosen randomly to give its state to the other. In node update dynamics (described above) a node is selected to adopt the state of a random neighbor. The neighbor chosen at random is likely to have high degree, so high degree nodes have more influence.

\section{Generalized voter model}

The two processes described above specify both the selection of a pair of interacting nodes and which node adopts the state of the other. We separate this process into two distinct steps to better understand the contribution of each aspect of the process.

Given a network of size $N$, each node $i$ has state $s_i = \pm 1$ and degree $k_i$. We define $P_{ij}$ to be the probability of giving to node $i$ the state of node $j$ during a given time step.  There are two independent components of this event: the probability $S_{ij}$ 
of selecting an edge connecting nodes $i$ and $j$ and the probability $W(k_i , k_j )$ that a node with degree $k_j$ gives its state to a node with degree $k_i$. Thus,
\begin{equation}
P_{ij} = S_{ij} W(k_i , k_j ).
\end{equation}
The form of $W (k_i , k_j )$ is motivated by comparison with node update dynamics 
in uncorrelated networks. In uncorrelated networks, node update dynamics can 
be described by considering all nodes of like degree to be indistinguishable in the ensemble average~\cite{Sood:2005p116}. 
In the following, $\{i\}$ and $\{j\}$ refer to the subgroups of all nodes 
with degrees $k_i$ and $k_j$, respectively. Under the node update process, the probability of giving a node with degree $k_i$ 
the state of a node with degree $k_j$ is 
\begin{equation}
P_{ij} = n_i \frac{k_j n_j}{\mu_1} 
\end{equation}
where $n_i$, $n_j$ are the fraction of nodes with degrees $k_i$ and $k_j$ and 
$\mu_1 = \sum_i n_i k_i$ is the average degree. This can be interpreted as the probability 
$n_i$ of selecting a node with degree $k_i$ times the probability of following an edge 
into a node with degree $k_j$. A node in $\{j\}$ has $k_j$ edges, so the probability of following an edge into it is proportional to $k_j$. If node pairs are selected in this manner, a particular pair of nodes in $\{i\}$ and $\{j\}$ can be chosen by either picking from $\{i\}$ and following an edge to a node in $\{j\}$ or by picking from $\{j\}$ and following an edge to a node in $\{i\}$. For node 
selection, this gives 
\begin{equation}
S^n_{ij} = \frac{n_i n_j}{2\mu_1} (k_i+k_j) .
\end{equation}
We propose a generalization that includes the standard node update dynamics, which requires $S_{ij} W(k_i,k_j) = P_{ij}$. The only form of $W(k_i,k_j)$ to do this is  
\begin{equation}
W (k_i , k_j ) = \frac{k_j}{k_i+k_j}.
\end{equation}
This form is also consistent with our definition of $W(k_i , k_j )$ as a probability. Notably, 
it suggests a one parameter generalization: 
\begin{equation}
W(k_i , k_j,\alpha) = \frac{k_j^\alpha}{k_i^\alpha + k_j^\alpha}.
\end{equation}

	Qualitatively, the parameter $\alpha$ determines how much a node asserts its degree when transmitting its state. 
For $\alpha > 0$, the higher degree node of a pair is more likely to give its state to the lower degree node, a bias that increases with $\alpha$. For $\alpha < 0$, the opposite is true. The special case $\alpha = 0$ ignores topology in determining the direction of interaction since $W(k_i,k_j,0) = 1/2$ always. Edge update dynamics is recovered by using edge selection to find pairs and setting $\alpha = 0$. Node update dynamics occurs when node selection determines pairs and $\alpha = 1$. A recently investigated ``invasion" dynamic, where a node is picked to give its state to a random neighbor (opposite the traditional model), occurs for node selection and $\alpha=-1$. If all nodes of the network have the same degree, as in a mean field or lattice topology, then all values of $\alpha$ are equivalent to the traditional voter model.

	 Effectively, this model assumes some connection between the behavior of the agents and the underlying network on which they live. For example, if this were to be thought of as a model of opinion dynamics, a value of $\alpha > 1$ under node update selection would correspond to a situation where an individual prefers to behave like those who are more connected than himself. Celebrities or well-regarded experts are extremely influential, for example, but the same forces that drive their high visibility also keep them from being influenced by the non-famous. The forces of influence and accessibility compete, such that a small value of $\alpha$ makes all nodes able to change state quickly, but limits the influence of any node. A high value of $\alpha$ makes high degree nodes influential, thus able to order their neighborhood quickly, but those influential nodes will flip only on rare occasions.

\subsection{Dynamics}
To understand the dynamics, we study the master equation for an arbitrary network. The probability of a system being in state $\mathbf{s} = \{s_i\}$ at time $t$ is defined to be $P(\mathbf{s}, t)$. Denote by $\mathbf{s}^i$ the state $\mathbf{s}$ where $s_i \mapsto -s_i$ and let $S_{ij}$ be the probability of selecting the edge between nodes $i$ and $j$. For brevity, we write $W(k_i , k_j, \alpha ) = W_{ij}$. The master equation is
\begin{equation}
\frac{d}{dt}{P(\mathbf{s},t)} = \sum_{ij}  S_{ij} W_{ij} \left( \frac{1+s_i s_j}{2} \right) P(\mathbf{s}^i,t)  - S_{ij} W_{ij} \left( \frac{1-s_i s_j}{2} \right) P(\mathbf{s},t)
\end{equation}
Let $\rho_k$ be the ensemble average probability of a node with degree $k$ being in a $+1$ state. Averaged over all random graphs, nodes of like degree are indistinguishable and we consider each such like-degree subgraph separately~\cite{Sood:2005p116}. The evolution of arbitrary ensemble average functions can be found in a straight-forward manner~\cite{Krapivsky92a}, giving:
 \begin{equation}
\label{rhoevolution}
\frac{d\rho_k}{dt} =  \sum_l \frac{S_{kl}}{n_k} W_{kl}(\rho_l - \rho_k).
\end{equation}
We can find a conserved magnetization, $\rho^*$, by choosing coefficients $C_i$ such that 
\begin{equation}
\frac{d\rho^*}{dt} = \sum_k C_k \frac{d}{dt}\rho_k = 0.
\end{equation}
Since $S_{ij}$ is symmetric in $i$ and $j$, this can happen for arbitrary subgraph densities only if
\begin{equation}
C_k \frac{W_{kl}}{n_k}= C_l \frac{W_{lk}}{n_l}.
\end{equation}
This implies that $C_k \propto n_k k^\alpha$. Normalizing,
\begin{equation}
\label{conservedmag}
\rho^* = \frac{\sum_k n_k k^\alpha \rho_k}{\sum_k n_k k^\alpha} = \frac{1}{\mu_\alpha} \sum_k n_k k^\alpha \rho_k
\end{equation}
where $\mu_\alpha$ is the $\alpha^{th}$ moment of the degree distribution. Note that the ensemble 
conserved magnetism is independent of the process of selecting node pairs. 

\section{Bipartite Network}
The simplest degree homogeneous topology is the fully connected bipartite network. Such a network is given by two groups of nodes, group $A$ with size $a$ and group $B$ with size $b$. A node in group $A$ is connected to every node in group $B$, but none in group $A$. The degree of nodes in $A$, $k_A$, is the size of $B$, giving $k_A = b$ and similarly $k_B = a$. In this situation all edges are interchangeable, so there is no difference between the two selection processes. We 
will consider only the effect of $\alpha$. Let $\rho_a$ be the concentration of 
$+1$ opinions in $A$ and $\rho_b$ be the concentration of $+1$ opinions in $B$. In our model, the special value $\alpha = 1$ is equivalent to the case studied in~\cite{Sood:2005p116} on the same network and we follow a similar procedure, omitting details that can be found there. From Equation~(\ref{conservedmag}), the conserved magnetization is 
\begin{equation}
\label{bipartrhostar}
\rho^* = \frac{1}{a^{\alpha-1} + b^{\alpha-1}}\left(b^{\alpha-1} \rho_a + a^{\alpha-1}\rho_b\right)
\end{equation}
For any initial conditions, the ensemble average subgraph densities approaches $\rho^*$. If all nodes in $A$ start as $+1$ and all nodes in $B$ start as $-1$, then the probability of ending in the $+1$ state is
\begin{equation}
P_+ = \frac{b^{\alpha-1}}{b^{\alpha-1} + a^{\alpha-1}}
\end{equation}

The mean exit time $T_\alpha$ is given by the backward Komologorov equation~\cite{Redner07a}. $T_\alpha$ solves
\begin{equation}
\label{bipartexittimeeq}
-\frac{a^\alpha+b^\alpha}{a+b} = (\rho_a - \rho_b)(b^{\alpha-1}\partial_b - a^{\alpha-1}\partial_a) T +\\ \frac{1}{2} (\rho_a + \rho_b - 2 \rho_a \rho_b) ( a^{\alpha-2} \partial^2_a + b^{\alpha-2} \partial^2_b) T.
\end{equation}
where $\partial_a$ and $\partial_b$ are partial derivatives with respect to the initial subgraph densities.The first term describes convection, which brings the subgraph densities to some equal value, and the second term describes the diffusion of the network-wide state~\cite{Sood:2005p116}. The convective dynamics can be shown to be fast for all $\alpha$.

\begin{figure}[htbp]
\begin{center}
\scalebox{0.7}{\includegraphics{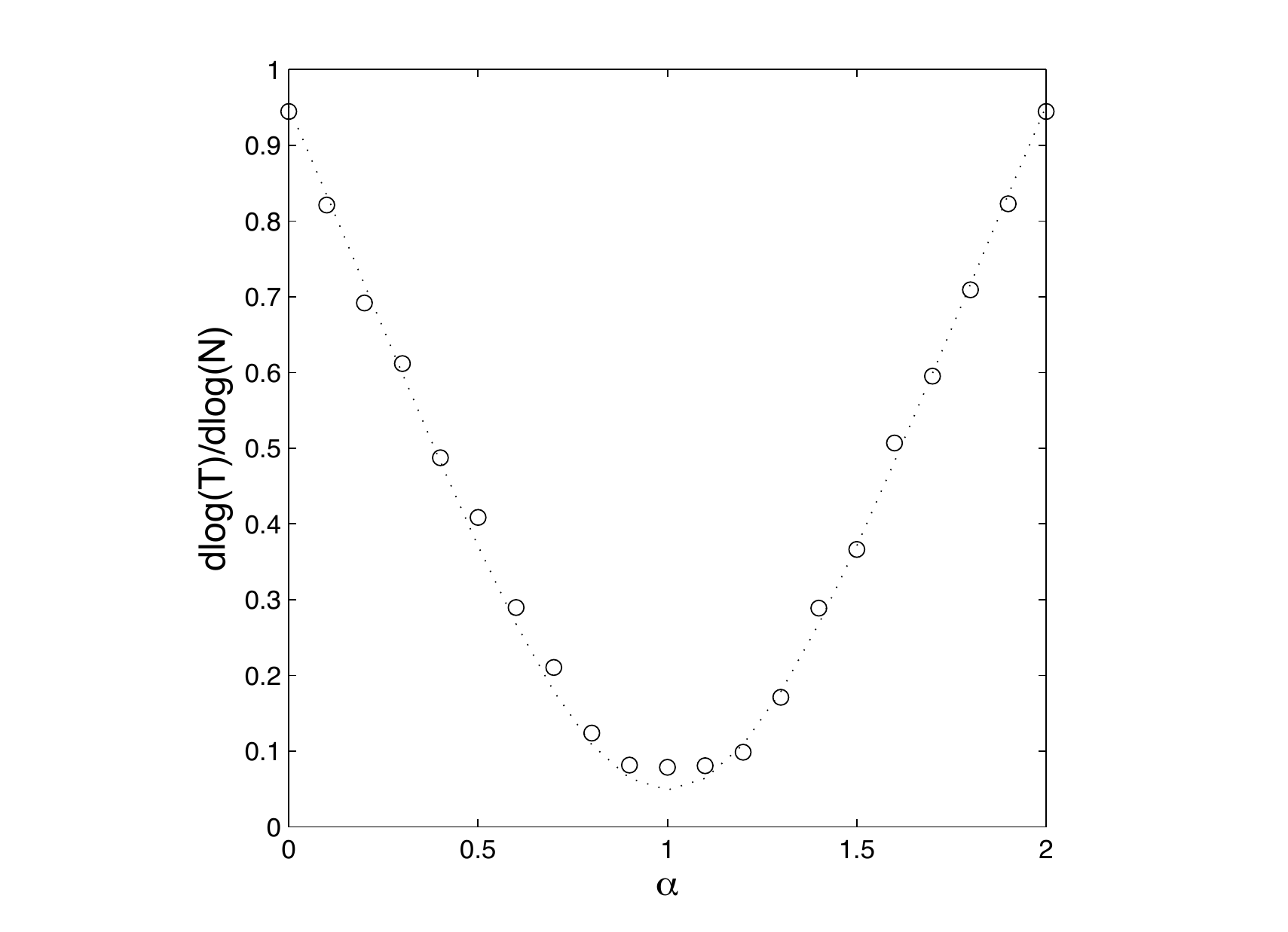}}
\caption{Circles are the simulated fit of $d \log{T}/ d \log{N}$ for a complete bipartite network with $M=40$ and $N$ ranging from 100 to 5000. The dotted line is the scaling of Equation~\ref{starScale} for the same range of $N$.}
\label{bipartScaling}
\end{center}
\end{figure}

The fast step toward equal subgraph densities has a negligible impact on extinction time and we can consider only the subsequent one dimensional problem. We define $\rho = \rho_a = \rho_b$ and apply a change of variables using Equation~(\ref{bipartrhostar}). After integrating,
\begin{equation}
T=-( a^{1-\alpha} + b^{1-\alpha} ) (a^{\alpha-1}+b^{\alpha-1})\frac{ab}{a+b} (\rho \log(\rho) + (1-\rho)\log(1-\rho)).
\end{equation}
This has a similar form to the standard voter model, but with a factor that is 
symmetric about $\alpha = 1$ and non-vanishing. If we take $a = \lambda b$, then 
\begin{equation}
T \propto (2 + \lambda^{1-\alpha} + \lambda^{\alpha-1}) \frac{\lambda}{1+\lambda} b.
\end{equation}
If $\lambda \gg 1$, corresponding to a star-like graph,
\begin{equation}
\label{starScale}
T \sim \lambda^{|\alpha-1|} b.
\end{equation}
This scaling is confirmed in simulations (see Figure~\ref{bipartScaling}). Notably, the standard voter model, $\alpha = 1$, is the fastest process for any complete bipartite network.

\section{Arbitrary networks}

Similar analysis extends naturally to networks in which a node's degree determines the network structure. Many random network models fall into this category, notably any random network generated by the configuration model, including those with scale-free distributions, and Erdos-Renyi networks~\cite{Newman:2003p90}. Small world networks are not included, however, as certain nodes have exceptional topological characteristics that are independent of their degree~\cite{Watts:1998p2416}.

As in Equation~(\ref{bipartexittimeeq}), we can write the equation satisfied by the mean exit time on an arbitrary network. In the following, $\rho_i$ refers to the density of $+1$ states on the subgraph of nodes with degree $k_i = i$.
\begin{equation}
\label{gentequation}
\delta_t = \sum_{ij} S_{ij}W^\alpha_{ij}(\rho_j-\rho_i)\delta_i\partial_i T + \frac{1}{2}\sum_{ij}S_{ij}W^\alpha_{ij}(\rho_i + \rho_j - 2\rho_i\rho_j) \delta_i^2 \partial_i^2 T
\end{equation}
The system is again split into a convective term and a diffusive term. The assumption of fast approach to well-mixed state must be treated more carefully in our generalized model, but there do exist cases where diffusion dominates. Namely, it was observed in~\cite{Sood:2005p116} that node update dynamics on a scale free network has fast convection compared to its diffusive exit time. There is, in general, a prefactor to the term $\rho_i(1-\rho_j)$ when evaluating the probability that two nodes have differing states \cite{Vazquez:2008p5347}. Numerical simulations show that this is of order one and not generally a constant over time when $\alpha \neq 1$, so it does not affect the final scaling results.

We expect this to not be true for all $\alpha$. In the case of $\alpha \gg 1$, a node with degree higher than all its 
neighbors will act to dictate its neighbors' states, but only rarely be changed itself. The network in this case may not be able to quickly approach the global equilibrium given by $\rho^*$, since these locally highest degree nodes will be pinned for a time dependent on $\alpha$. If this duration is longer than the time for the rest of the system to become ordered via mixing and diffusion, a quasi-frustrated state occurs. The exact local topology, rather than just degree distributions, can dominate the dynamics.

Let us suppose that the exit time is diffusion dominated and will return to discuss the validity of this assumption. The system can be approximated by a one dimensional equation in $\rho = \frac{1}{\mu_\alpha} \sum_i n_i k^\alpha_i \rho_i$:
\begin{equation}
-N = \frac{1}{\mu^2_\alpha} \left( \sum_{ij}S_{ij}W^\alpha_{ij}k_i^{2\alpha}\right) \rho (1 - \rho) \partial_\rho^2 T.
\end{equation}
And thus 
\begin{equation}
T \propto -\frac{N\mu^2_\alpha}{\sum_{ij}S_{ij}W_{ij}^\alpha k_i^{2 \alpha}}.
\end{equation}
The denominator can be simplified by noting that
\begin{equation}
\sum_{ij} S_{ij}W^\alpha_{ij}k_i^{2 \alpha} = \frac{1}{2} \left( \sum_{ij} S_{ij} W^\alpha_{ij} k_i^{2\alpha} + \sum_{ij} S_{ij} W^\alpha_{ji} k_j^{2\alpha} \right) = \frac{1}{2} \sum_{ij} S_{ij} k_i^\alpha k_j^\alpha.
\end{equation}
This gives:
\begin{equation}
\label{diffexitscale}
T \propto N \frac{\mu_\alpha^2}{\sum_{ij} S_{ij} k_i^\alpha k_j^\alpha}.
\end{equation}
Since $S_{ij}$ is a probability, the sum can be thought of as a weighted average over 
selection probabilities. Interestingly, for $\alpha = 0$, neither the form of interaction selection nor the network topology matter. In 
that case, $\mu_0 = 1$ and $W_{ij} = 1/2$, so 
\begin{equation}
T_{\alpha=0} \propto N.
\end{equation}
This agrees with the observation in~\cite{Suchecki05a} that exit times scale with $N$ in situations where the unweighted magnetization is conserved, which corresponds exactly with $\alpha = 0$.

\begin{figure}[htbp]
\begin{center}
\scalebox{0.7}{\includegraphics{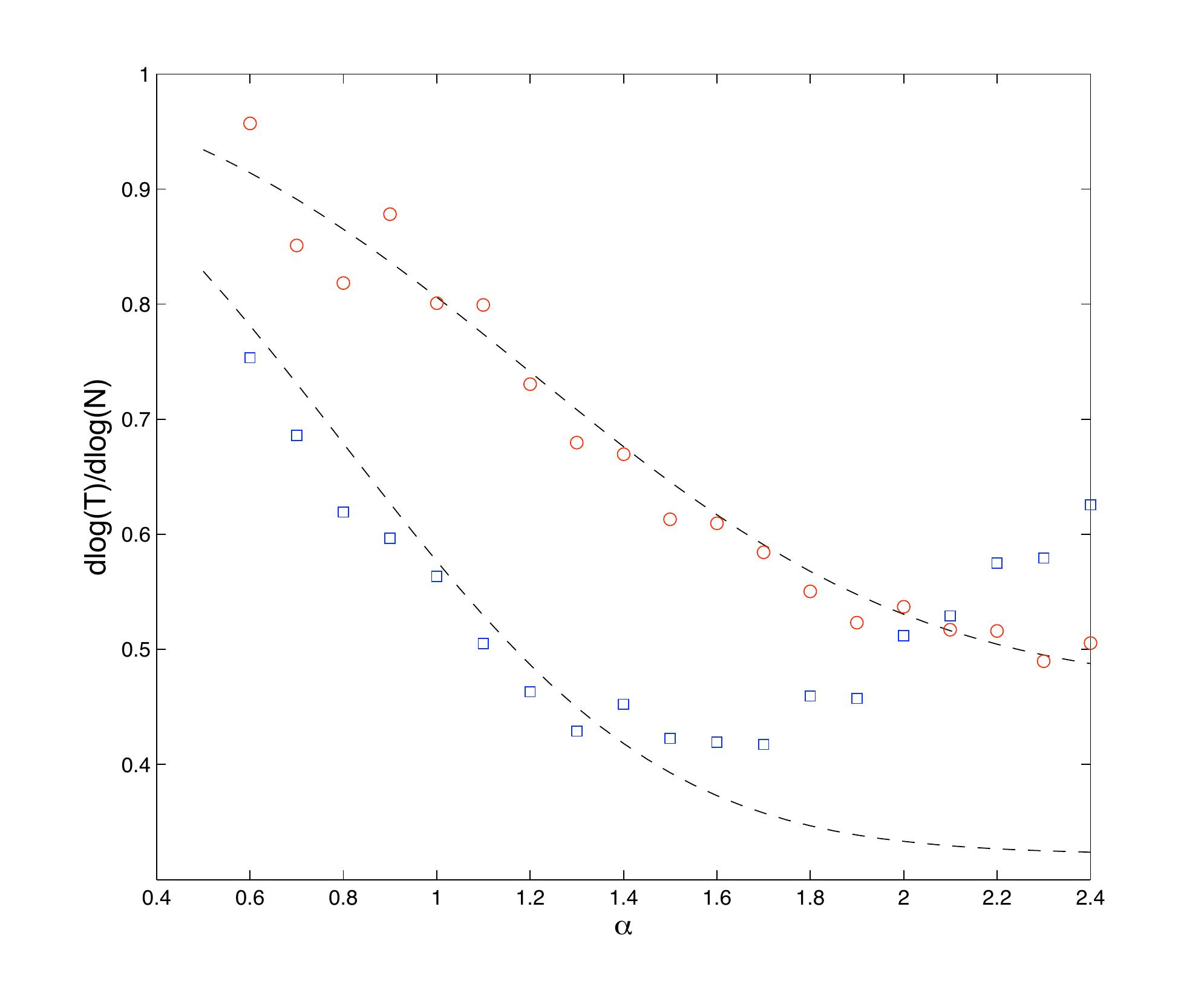}}
\caption{Simulated values of $d \log{T}/ d \log{N}$ under \emph{node} selection based on several hundred runs for $N$ from 750--15000. Circles correspond to $\nu = 2.8$, squares to $\nu=2.4$. The dashed line is the scaling based on the diffusive estimate, calculated by fitting the numerically calculated sum in Equation~\ref{diffexitscale} for similar values of $N$. Note that frustration begins to dominate for $\nu=2.4$ at $\alpha>1.6$, causing the deviation from the diffusive estimate.}
\label{nodeScaling}
\end{center}
\end{figure}

To go farther, we need to specify the selection scheme and the network. We focus our consideration on random uncorrelated scale-free networks with degree distribution $n_k \sim k^{-\nu}$. Networks with power law distributions appear in a variety of social and biological contexts and exhibit a range of interesting behaviors~\cite{Newman:2003p90}. Let us first consider node update, for which $S_{ij} = n_i n_j \frac{k_i + k_j}{2\mu_1}$.
Then Equation~(\ref{diffexitscale}) becomes:
\begin{equation}
T_\mathcal{N} \propto N \frac{\mu_1 \mu_\alpha}{\mu_{\alpha+1}}.
\end{equation}
The $\alpha^{th}$ moment can be approximated by an integral:
\begin{equation}
\mu_\alpha \sim \int^{k_{max}} k^\alpha n(k) dk
\end{equation}
 up to an effective maximum degree $k_{max}$, defined by $\int_{k_{max}}^\infty n(k) dk = 1/N$~\cite{Krapivsky:2002p57}. It is easily seen that $k_{max} \sim N^{1/(\nu-1)}$. 
\begin{equation}
T_{\mathcal{N}} \propto \left\{ \begin{array}{ll}
N^{\frac{\nu-2}{\nu-1}} & \alpha > \nu - 1 \\
N^{\frac{2 \nu - \alpha - 3}{\nu - 1}} & \nu - 2 < \alpha < \nu - 1 \\
N & \alpha < \nu - 2 \\
\end{array}
\right.
\end{equation}
For $\nu > 2$ and any $\alpha$, the exit time increases without bound as system size increases. We simulated the process on random network generated by the configuration model~\cite{Molloy:1995p2460} and found good agreement with our predictions (see Figure~\ref{nodeScaling}).

\begin{figure}[htbp]
\begin{center}
\scalebox{0.7}{\includegraphics{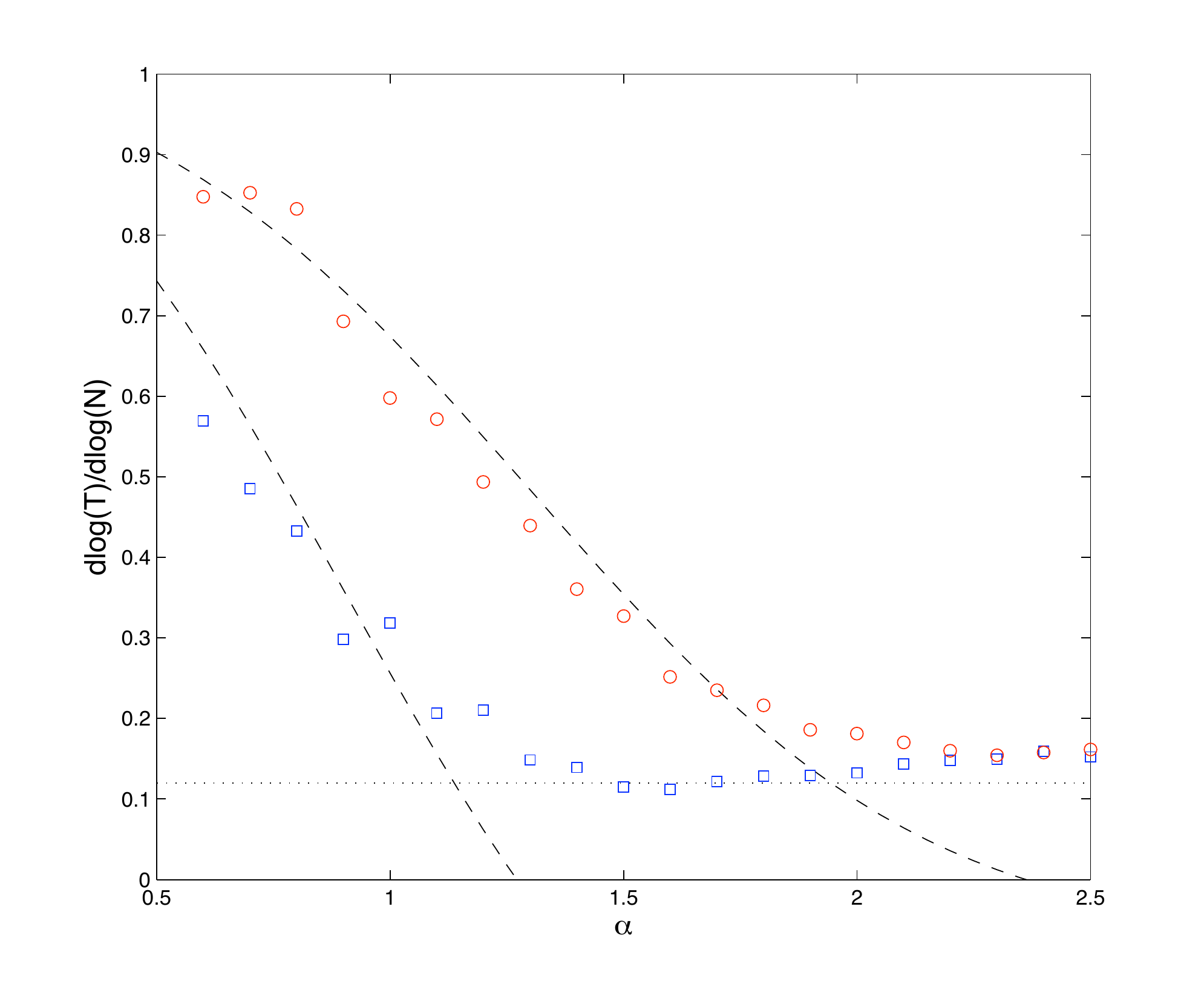}}
\caption{Simulated values of $d \log{T}/ d \log{N}$ under \emph{edge} selection from several hundred runs of values of $N$ from 750--15000. Circles are for $\nu = 2.8$, squares for $\nu=2.4$. The dashed line comes from fitting the numerically calculated sum in Equation~\ref{diffexitscale} for similar values of $N$. The horizontal line is the effective slope of $T \sim \log{N}$ for the system sizes used.}
\label{edgeScaling}
\end{center}
\end{figure}

For edge update dynamics, $S_{ij} = n_i n_j \frac{k_i  k_j}{\mu^2}.$
Low degree nodes are selected less frequently under edge selection than node selection. The diffusive exit time can be calculated similarly, giving:
\begin{equation}
T_\mathcal{E} = N \left(\frac{\mu_1 \mu_\alpha}{\mu_{\alpha+1}}\right)^2.
\end{equation}
The approximate scaling for edge update is
\begin{equation}
T_\mathcal{E} \propto \left\{ \begin{array}{ll}
N^{\frac{\nu-3}{\nu-1}} & \alpha > \nu - 1 \\
N^{\frac{3 \nu - 2\alpha - 5}{\nu - 1}} & \nu - 2 < \alpha < \nu - 1 \\
N & \alpha < \nu - 2 \\
\end{array}
\right.
\end{equation}
This leads to very different scaling behavior. For the parameter regions
\begin{align}
\nu < 3 & \quad \quad  \alpha > \nu - 1 \\
\nu < \frac{2\alpha + 5}{3} & \quad \quad   \nu - 2 < \alpha < \nu - 1
\end{align}
the diffusive exit time vanishes as $N$ increases. Simulations show that there is a diffusive region which agrees with our predictions for smaller values of $\alpha$ (see Figure~\ref{edgeScaling}).

The convective process involves an exponential decay of each $\rho_i$ to its equilibrium value, with rate determined by the network structure. When this value is comparable to $1/n_i N$ (or $1 - 1/n_iN$) for all $i$, the system effectively reaches convergence before diffusive time scales matter. We can read off the dynamics  from Equation~\ref{rhoevolution}. The decay rate $\tau$ is not strongly dependent on system size, so the relationship with $N$ comes only from the exit condition
\begin{equation}
\frac{1}{N} \sim e^{-T_\mathcal{E}/\tau}.
\end{equation}
This implies that when diffusive time scales vanish,
\begin{equation}
T_\mathcal{E} \sim \log(N).
\end{equation}
Exponential mixing is only observed as a dominant time scale under link update selection for values of $N$, $\nu$, and $\alpha$ such that frustration does not occur.

\begin{figure}[htbp]
\begin{center}
\scalebox{0.6}{\includegraphics{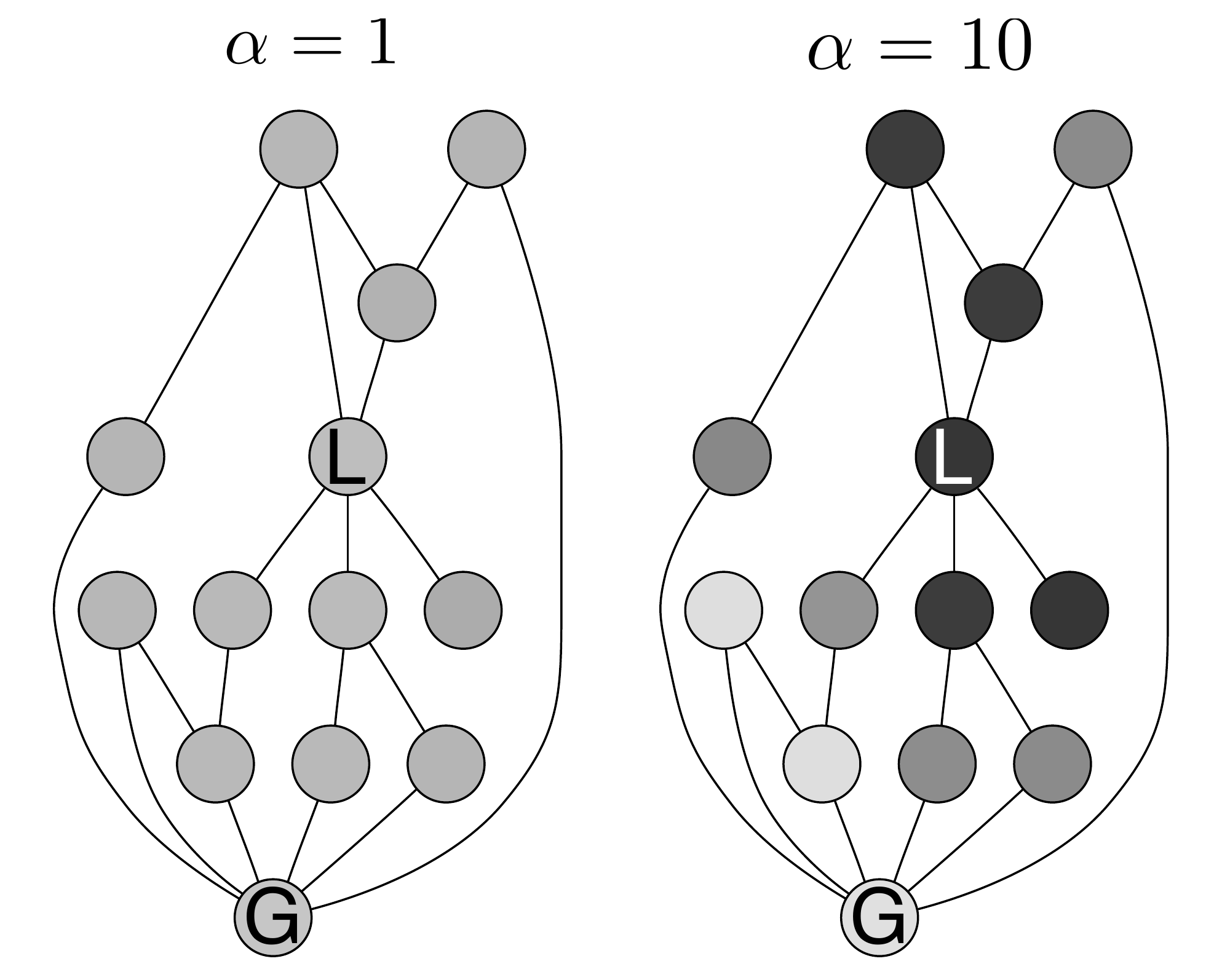}}
\caption{A example of node update dynamics on a small network containing a global highest degree node ($G$) and a separate local highest degree node ($L$) for $\alpha=1$ and $\alpha=10$. The darkness of a node corresponds to the average fraction of time spent in a state opposite the final state of the network over 1000 realizations with identical initial conditions. A darker node has spent more time in a contrary state than a light node. For $\alpha=1$, states are well mixed and $T=9.4$. For $\alpha=10$, mixing does not occur. The local leader and its neighborhood spend most of the time in a contrary state and $T=4814$.}
\label{netFrust}
\end{center}
\end{figure}

To understand where frustration comes from, consider a node whose degree is higher than that of all of its neighbors. In the limit $\alpha \rightarrow \infty$, such a node will take an arbitrarily long time to flip. For large values of $\alpha$, escape from this frustrated state must then be the dominant time scale, as diffusive time scales for high $\alpha$ are dependent only on $\nu$. For either selection process, the probability of a given node being sampled in a time step is a constant, independent of system size. It is also possible for small connected subgraphs to be frustrated if every edge out of the subgraph is to a node of lower degree. As a result of locally highest degree nodes and clusters, a small number of hard-to-flip nodes can hold back a large network from reaching convergence (Figure~\ref{netFrust}).

\begin{figure}[htbp]
\begin{center}
\scalebox{0.8}{\includegraphics{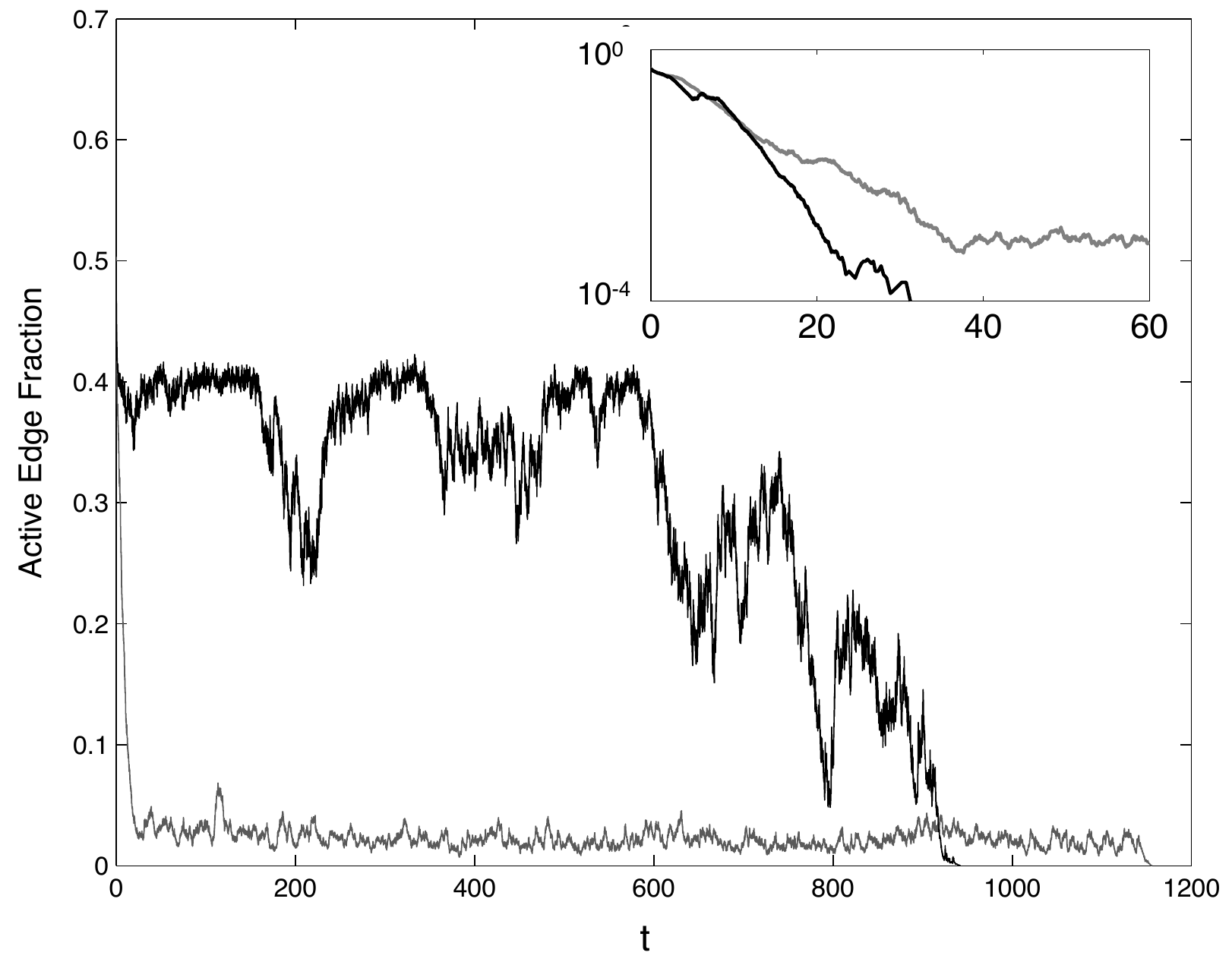}}
\caption{Typical dynamics of the fraction of active edges (edges connecting nodes with different states) for diffusion and frustration. Both have $N=10000$, $\nu=2.8$, and edge selection. The top line, in black, is for $\alpha=1$ and the bottom line, in gray, is for $\alpha=8$. The $\alpha=1$ case is diffusive and fluctuates to convergence after reaching the ensemble average value. The $\alpha=8$ case decays exponentially to a value greater than the ensemble average value, because a locally highest degree node or cluster is slow to flip. Just before $t=1200$, these nodes flip and the system reaches the absorbing state. Inset shows early time dynamics on a similar network for $\alpha = 10$ in gray and $\alpha=4$ in black. Both decay exponentially initially, but the high $\alpha$ case becomes frustrated and the other continues to convergence.}
\label{dynamicsPlot}
\end{center}
\end{figure}

The continuum treatment in Equation~\ref{gentequation} averages over network ensembles before solving for the exit time. In the high $\alpha$ case, considering an averaged network ignores local leader effects and fails to give an accurate solution. The approach to an ensemble equilibrium state does not occur, so diffusion about this state is not a valid assumption (Figure~\ref{dynamicsPlot}).

The degree distribution of locally highest degree nodes can be approximated for the non-assortative case quite simply. The degree distribution of local leaders, $p_{ll}(k)$, is the independent product that a node has degree $k$ and that all $k$ neighbors have a degree less than $k$~\cite{Blondel07a}:
\begin{equation}
\label{locleaderdist}
p_{ll}(k) = p(k) \left( \sum_{k'<k} \frac{k' p(k')}{\mu_1} \right)^k.
\end{equation}
No term is strongly related to system size, so the total number of local leaders $N_{ll}$ scales linearly with $N$. 

The dynamics of these nodes are based on extremely local behavior and thus very hard to approximate, but we can construct the slowest possible node and understand its behavior. For a sufficiently large system, one such node will likely exist and thus dominate the exit time. The probability on any given time step to flip a local leader with degree $l$ is formally given by:
\begin{equation}
P_{flip}(l) = \sum_i S_{il} \frac{k_i^\alpha}{k_i^\alpha+l^\alpha}  P(\sigma_i \neq \sigma_{l})
\end{equation}
where $P(\sigma_i \neq \sigma_l)$ is the probability that the neighboring state differs from the local leader's state and subscripts index individual nodes. 

For either selection process, when $l$ is large,
\begin{equation}
S_{li} \sim \frac{1}{N}.
\end{equation}
We assume the probability of a neighbor being in the opposite state is finite and treat the following as a lower bound. Simulations show that this is reasonable. Assuming that $l^\alpha \gg k_i^\alpha$,
\begin{equation}
P(l) \approx \frac{1}{N l^{\alpha -1}} \left( \frac{1}{l}\sum_i k_i^\alpha \right).
\end{equation}
The slowest situation occurs when the local leader in question is surrounded by relatively low degree nodes and $P(l) \sim l^{1-\alpha}/N$. For this case, the time to flip in units of system size scales as:
\begin{equation}
T_f \sim l^{\alpha-1}.
\end{equation}
The highest degree of the local leaders scales linearly with the global highest degree ~\cite{Blondel07a}, so for $\nu<3$ we have $T_f \sim N^{(\alpha-1)/(\nu-1)}$. This will be true for large $\alpha$, but requires the system size to be high enough that at least one rare node such as the one described is likely. Such networks are prohibitively large to sample, but the qualitative situation described explains why frustration sets in at a smaller value of $\alpha$ for $\nu=2.4$ than $\nu=2.8$ in Figure~\ref{nodeScaling}. Local leaders for small $\nu$ have a higher degree than for large $\nu$, though numerical comparison from Equation~\ref{locleaderdist} shows them to be rarer.

\section{Summary}

Recent work~\cite{Baxter:2008p441} has found an approximate mean exit time for a duplication process on networks with arbitrary edge weights, assuming that diffusion is the dominant time scale. In this work, we demonstrate that there are at least two natural ways for this estimate of exit time to fail. As observed in Baxter \emph{et al.}, the time for the system to reach a metastable equilibrium can be at least as large as the diffusive exit time scale. We see this in the edge selection process for values of $\alpha$ and $\nu$ where the diffusive exit time vanishes as system size gets large. The frustrated dynamics in the node selection mode, however, presents a new way in which the diffusive estimate can fail. System dynamics are driven by a small number of topologically special nodes, breaking the assumption that a continuum description applies.

We have defined and analyzed a single parameter voter-model-like stochastic process that is identical to the original voter model on a homogeneous network. On a complex network, our generalized voter model has a tunable dependence on local network topology, allowing us to control the differing effects of selection and the direction of influence. On complete bipartite graphs, the traditional voter model is the fastest process to reach an absorbing state. On scale free networks, the dynamics depend strongly on the selection process. Node selection has two regimes; a diffusive one characterized by a well defined average magnetism and diffusion constant based on global network properties, and one with frustrated dynamics stemming from the neighborhood around a small number of locally highest degree nodes. Edge selection, previously considered uninteresting, has three regimes. In addition to diffusive and frustrated regimes, it also has a middle ground characterized by exponential mixing. Understanding the dynamics involved in this transition to frustration would be an interesting avenue for future work.

\section{Acknowledgements}

LMS is supported in part by NSF grant DMS 0554587. CMS would like to thank A. Stein, S. Cobey, and D. Adams for useful conversations.

\bibliography{allbiblio.bib}

\end{document}